\documentclass[12pt,a4paper]{article}
\usepackage{amsfonts}
\usepackage{amssymb}
\usepackage{indentfirst}
\usepackage{amsmath}
\usepackage{epsfig}
\usepackage{graphicx}
\usepackage[numbers,sort&compress]{natbib}
\usepackage{subfig}
\usepackage[symbol]{footmisc}
\usepackage{rotating}
\usepackage[dvipsnames]{xcolor}
\usepackage{booktabs}
\usepackage{caption}
\usepackage{array}
\usepackage{dutchcal}
\usepackage[dvipdfm,colorlinks,linkcolor=blue,citecolor=blue,urlcolor=blue]{hyperref}
\usepackage{url}

\makeatletter

\long\def\@makecaption#1#2{%
\vskip\abovecaptionskip
\sbox\@tempboxa{{\bfseries#1.}\hskip 1em#2}%
\ifdim \wd\@tempboxa >\hsize
   {\bfseries#1.}\hskip 1em#2\par
\else
 \global \@minipagefalse
 \hb@xt@\hsize{\hfil\box\@tempboxa\hfil}%
\fi \vskip\belowcaptionskip} \makeatother
\newcommand{\rA}{\tilde{r}_{\text{A}}}

\newcommand{\rE}{\tilde{r}_{\text{E}}}
\newcommand{\drA}{\dot{\tilde{r}}_{\text{A}}}
\newcommand{\drE}{\dot{\tilde{r}}_{\text{E}}}
\newcommand{\R}{\tilde{r}}
\newcommand{\dR}{\dot{\tilde{r}}}

\begin{document}
\title{\textbf{Thermodynamics of arbitrary spherical surfaces in FLRW spacetimes}}
\author {\small {Xun Gong, Jun Wang\footnote{\href{mailto:wjun@ynu.edu.cn}{E-mail:
wjun@ynu.edu.cn}}}, Chen-Yang Zhang}
\date{{\small{\centerline{\ School of Physics and Astronomy,
Yunnan University, Kunming 650500, P. R. China}}}}
\maketitle
\vspace{-2cm}
\begin{abstract}
 In this work, we investigate the thermodynamics of arbitrary spherically symmetric surfaces in non-flat FLRW spacetimes within general relativity. Adopting the gravity-to-thermodynamics perspective, the Clausius relation and the unified first law follow directly from the gravitational field equations and matter dynamics, with a central role played by the matter-flux measure derived from energy-momentum conservation. The cosmological apparent-horizon and event-horizon thermodynamics are both recovered as special cases. In particular, the event-horizon temperatures derived from the gravitational dynamics satisfy both thermodynamic relations without introducing phenomenological modifications, and the corresponding temperature reduces to the Gibbons-Hawking temperature in the de Sitter limit. We further show that previous inconsistencies originate from incompatible temperature prescriptions rather than from any intrinsic deficiency in the thermodynamics of the cosmological event horizon. These results demonstrate that the gravity-thermodynamics correspondence is a generic feature of gravitational dynamics rather than a property unique to causal horizons.


\begin{flushleft}
\sl {PACS~: 98.80.-k, 98.80.Jk, 04.20.-q}
\end{flushleft}
\end{abstract}

\section{Introduction}

 The relationship between gravity and thermodynamics was first recognized through the formulation of black hole thermodynamics~\cite{Bekenstein1972,Bekenstein1973,Bekenstein1974,BCH1973,Hawking1975,Hawking1976}, which demonstrated that a black hole event horizon possesses an entropy and a temperature proportional to its area and surface gravity, respectively, and obey laws analogous to those of ordinary thermodynamics. This entropy--area relation subsequently led 't~Hooft~\cite{tHooft1993} and Susskind~\cite{Susskind1995} to formulate the holographic principle, which asserts that the degrees of freedom of a gravitational system reside on its boundary rather than in the bulk~\cite{Bousso2002}. Notably, such a holographic structure is already encoded at the classical level in the Einstein--Hilbert Lagrangian through its intrinsic holographic relation~\cite{Padmanabhan2005,Padmanabhan2010new,Padmanabhan2013,Khodabakhshi2022}.

 In dynamical black hole spacetimes, however, the event horizon is not locally defined. To overcome this limitation, Hayward formulated a thermodynamic framework based on the apparent horizon, known as the unified first law ~\cite{Hayward1998}, in which the temperature is associated with the Kodama surface gravity~\cite{Hayward1998,Helou2015}.
 Further developments along this direction can be found in Refs.~\cite{AshtekarKrishnan2003,BoothFairhurst2004,NielsenVisser2006,AshtekarKrishnan2004review,Wald2001review,Hollands2024,Ashtekar2025}.
 An even more remarkable insight was provided by Jacobson, who reversed the conventional perspective and derived the Einstein field equations as an equation of state from the Clausius relation ($TdS=\delta Q$) at a local Rindler horizon, where $\delta Q$ and $T$ represent the energy flux and the Unruh temperature, respectively~\cite{Jacobson1995}. These results point to a profound connection between gravity and thermodynamics and suggest that gravitation may possess both holographic and emergent features.
 Present studies have largely proceeded along two complementary approaches: deriving thermodynamic laws from gravitational dynamics and deriving gravitational dynamics from thermodynamic principles. In either framework, thermodynamic quantities are expressed in geometric terms.

 The thermodynamic ideas originally developed in black hole physics can be naturally generalized to cosmology. Since the FLRW spacetime is intrinsically dynamical, it faces the same challenge as dynamical black holes, namely, the absence of a locally defined event horizon. Consequently, thermodynamic investigations have primarily focused on the apparent horizon. In 2005, Cai and Kim~\cite{Cai2005} generalized Jacobson's thermodynamic derivation of gravitational dynamics to the cosmological apparent horizon and showed that the Clausius relation, $T_{\text{CK}}dS = \delta Q$, reproduces the Friedmann equations, where $T_{\text{CK}}=1/(2\pi \rA)$ is termed the Cai--Kim temperature in Ref.~\cite{Tian2015} and $\rA$ denotes the apparent-horizon proper radius~\cite{Cai2005}. Subsequently, Cai et al.\ showed that the Friedmann equations can be recast into the thermodynamic form~\cite{CaiCao2007}, i.e., $T_{\text{KH}} dS = -dE + W dV$, where $T_{\text{KH}}$ is proportional to the Kodama surface gravity and is referred to the Kodama--Hayward temperature~\cite{Tian2015}, while $W$ denotes the work density~\cite{CaiCao2007}. This relation is also known as the unified first law~\cite{AkbarCai2007}, as it is formally identical to the unified first law of dynamical black hole thermodynamics. Moreover, thermodynamics associated with the apparent horizon has been successfully established not only in general relativity but also in a wide range of modified theories of gravity (see Ref.~\cite{Wang2020} and references therein). Although the Clausius relation and the unified first law share the same entropy expression, they rely on different notions of temperature. Nevertheless, several studies have uncovered nontrivial connections between these temperature definitions. A notable example is that the Kodama--Hayward temperature approaches the Cai--Kim temperature in the limit $\drA \to 0$~\cite{Cai2005,CaiCao2007,AkbarCai2007}.

 In contrast to the success of thermodynamics on apparent horizon, establishing a consistent thermodynamic description for the cosmological event horizon has proven considerably more challenging. As early as 1977, Gibbons and Hawking investigated the thermodynamics on the event horizon in de Sitter spacetime and showed that it is characterized by an entropy proportional to the area of the event horizon and a temperature $T_{\text{GH}}=1/(2\pi \rE)$, known as the Gibbons--Hawking temperature, where $\rE$ denotes the proper radius of the event horizon~\cite{gibbons_cosmological_1977}.
 Subsequently, it was demonstrated that these quantities obey the corresponding thermodynamic laws  in Ref.~\cite{Frolov_2003}.  However, extending this approach to a general cosmological event horizon has encountered significant difficulties. Wang et al.\ showed that neither the Gibbons--Hawking temperature nor any simple rescaling thereof satisfies the Clausius relation~\cite{wang_thermodynamics_2006}, which indicates that the cosmological event horizon may not constitute a genuine thermodynamic boundary. Subsequently, Chakraborty demonstrated that, in a spatially flat FLRW universe, the redefined temperature $T=H^2\rE /2\pi$ allows the Clausius relation to hold for static dark energy models~\cite{chakraborty_is_2012}, although it fails in holographic dark energy scenarios. Later, a more general temperature prescription $T=\alpha \rE/(2\pi \rA^2)$, where $\alpha = (\drA/\rA)/(\drE/\rE)$, was proposed, under which the Clausius relation can be recovered on the event horizon~\cite{Chakraborty2014,Haldar2017}. The validity of this prescription in $f(R)$ theories of gravity has also been confirmed in Ref.~\cite{tu_general_2016}. However, Gohar and Salzano recently argued that such temperature rescaling procedures do not consistently implement the underlying thermodynamic relations~\cite{GOHAR2024EntropicCosmology}, underscoring the lack of consensus regarding the appropriate temperature associated with the cosmological event horizon. It is worth emphasizing that all of the aforementioned studies attempt to restore the Clausius relation through temperature redefinitions, leading to temperature expressions that differ from the original Gibbons--Hawking temperature.

 Thermodynamic studies of black hole and cosmological horizons have provided compelling evidence that gravity may not be fundamental but instead emerges from more microscopic degrees of freedom, much like the macroscopic behavior of fluids emerges from molecular dynamics. This perspective is embodied in the emergent gravity paradigm~\cite{Padmanabhan2010new,Padmanabhan2015}, in which the holographic principle provides a natural explanation for the entropy--area law of horizons, while Padmanabhan's ``heating spacetimes'' interpretation offers an intuitive physical picture for gravitational dynamics~\cite{Padmanabhan2015,Padmanabhan2008,padmanabhan2009,Padmanabhan2010}. In this framework, phenomena such as the Unruh effect and Hawking radiation imply that spacetime can be heated. Motivated by Boltzmann's insight that any system capable of being heated must possess microscopic degrees of freedom, Padmanabhan proposed that spacetime is composed of a vast number of ``atoms of spacetime'', with gravitational dynamics emerging as a macroscopic manifestation of their collective behavior~\cite{padmanabhan2009,Padmanabhan2010new,Padmanabhan2012}. These ideas further inspired Verlinde to propose an alternative emergent gravity scenario in which gravity arises as an entropic force associated with entropy variations on holographic screens~\cite{Verlinde2011}. Such screens are codimension-one hypersurfaces that generalize the notion of horizons to generic spacetimes and exhibit rich thermodynamic properties~\cite{Chatwin-Davies2024}. Subsequently, the thermodynamics of holographic screens in static spacetimes was systematically explored in Refs.~\cite{Tian2011,Chen2011}. More recently, the thermodynamic properties of a class of spacetimes with no horizons but containing specific timelike junctions were investigated in Ref.~\cite{Isichei2025withoutH}. These developments indicate that the connection between gravity and thermodynamics is not restricted to horizons but can also be established on non-horizon hypersurfaces, such as holographic screens, in static spacetimes.

 For dynamical spacetimes, however, it remains unclear whether the connection between gravity and thermodynamics extends beyond horizons to more general surfaces. To explore this possibility, we investigate thermodynamics of arbitrary spherically symmetric surfaces in non-flat FLRW spacetimes and subsequently apply the general formalism to both the cosmological apparent horizon and the event horizon.

 The remainder of this paper is organized as follows. In Sect.~\ref{sec:TASB}, we develop a thermodynamic framework for arbitrary spherically symmetric surfaces in FLRW spacetimes. In Sect.~\ref{sec:TAH}, we specialize the obtained formalism to the cosmological apparent horizon and recover the standard thermodynamic relations, including the Clausius relation and the unified first law. Following this, Sect.~\ref{sec:APP} delves into the applications of our general formulas. In Sect.~\ref{sec:TEH}, we apply the framework to the cosmological event horizon and derive a self-consistent horizon temperature without introducing modified temperature prescriptions. This naturally leads to both the Clausius relation and the unified first law on the event horizon, thereby resolving the longstanding ambiguities surrounding its thermodynamic description. Finally, Sect.~\ref{sec:Conclusion} presents our conclusions. Unless otherwise stated, all calculations are performed in four-dimensional spacetime with metric signature $(-,+,+,+)$ and adopt units $\hbar=c=k_{\text{B}}=1$.

\section{Thermodynamics of arbitrary spherically symmetric surfaces}
\label{sec:TASB}

According to Padmanabhan, spacetime can be ``heated'' through two distinct mechanisms~\cite{padmanabhan2009}. One arises from the gravitational collapse of matter, which gives rise to a black-hole horizon endowed with a well-defined Hawking temperature. The other originates from the acceleration of an observer, which generates a local Rindler horizon associated with the Unruh temperature. In both cases, the dynamical behavior of matter around a horizon endows the horizon with genuine thermodynamic characteristics, thereby enabling it to exchange energy with its surroundings~\cite{Padmanabhan2010Interpretation}.

Once a horizon acquires thermodynamic attributes, a virtual displacement of the horizon along its normal direction establishes a direct connection to gravitational dynamics. If such an infinitesimal displacement engulfs a small amount of energy flux, $\delta Q$, located just outside the horizon, consistency with the thermodynamic laws requires the horizon entropy to vary according to the Clausius relation, $\delta Q = T \delta S$, where $T$ denotes the horizon temperature. Remarkably, imposing this relation on all local Rindler horizons leads directly to the Einstein field equations~\cite{Jacobson1995}. In this sense, gravitational dynamics emerges from the thermodynamic requirement that horizons obey the fundamental laws of thermodynamics, with horizons serving as the essential bridge between thermodynamics and gravity.

Since horizons represent only a special class of distinguished surfaces, it is natural to ask whether the thermodynamic description of spacetime can be formulated in terms of general spacetime surfaces, rather than being restricted to horizons alone. In this section, we investigate the connection between thermodynamics and cosmological dynamics on arbitrary spherically symmetric surfaces in non-flat FLRW spacetimes within the framework of general relativity. To this end, one must first establish that arbitrary spherically symmetric surfaces can be endowed with well-defined thermodynamic properties.
\subsection{Matter-flux measure and thermodynamic interpretation}

 To establish the thermodynamic properties of arbitrary spherically symmetric surfaces, we start from the covariant conservation law of the matter energy-momentum tensor,
 \begin{equation}
 \label{CDTM}
   \nabla^{\mu} T_{\mu\nu}=0,
 \end{equation}
  which may be regarded, in a certain sense, as the equation of motion for matter~\cite{misner1973gravitation,BEKENSTEIN2015337}. It therefore encodes the dynamical behavior of matter around a horizon or, more generally, an arbitrary surface.

 To proceed, we consider a non-flat FLRW spacetime with the line element,
\begin{align}
\label{FLRW}
ds^2=-dt^2+a^2(t)\left[\frac{dr^2}{1-kr^2}
+r^2(d\theta^2+\sin^2\theta d\phi^2)\right],
\end{align}
where $a(t)$ is the scale factor and $k\in \{-1,0,+1\}$ denotes the spatial curvature parameter corresponding to open, flat, and closed spatial geometries, respectively. According to the cosmological principle, the cosmic medium can be considered as a perfect fluid~\cite{misner1973gravitation},
\begin{align}
\label{fluidT}
T_{\mu\nu}=(\rho+P)u_{\mu}u_{\nu}+Pg_{\mu\nu},
\end{align}
where $\rho$ and $P$ are the energy density and pressure, respectively, and $u_{\mu}=(-1,0,0,0)$ is the four-velocity of the fluid.

Substituting Eqs.~\eqref{FLRW} and \eqref{fluidT} into the Einstein field equations yields the Friedmann equations,
\begin{align}
H^2+\frac{k}{a^2}
&=\frac{8\pi G}{3}\rho,
\label{F1}
\\
\dot H-\frac{k}{a^2}
&=-4\pi G(\rho+P),
\label{F2}
\end{align}
where an overdot denotes differentiation with respect to cosmic time.
Unlike the static spherically symmetric case, where the Einstein equations can be rewritten as thermodynamic identities on arbitrary holographic screens~\cite{Tian2011,Chen2011}, Eqs.~\eqref{F1} and~\eqref{F2} explicitly depend on time. Consequently, the thermodynamic interpretation of gravity in dynamical cosmological spacetimes is far from straightforward.

In FLRW spacetime, the proper radius of an arbitrary spherically symmetric surface can be expressed as
\begin{equation}
\label{ASB}
\R(t)=a(t)\,\mathcal{r}(t),
\end{equation}
where the function $\mathcal{r}(t)$ denotes the radial coordinate of the surface. This provides a unified description of all spherically symmetric surfaces without imposing any additional geometric or causal conditions. The two cosmological horizons most commonly studied in the literature can be recovered by imposing different constraints on $\mathcal{r}(t)$. When the surface is required to be marginally trapped, Eq.~\eqref{ASB} reduces to the cosmological apparent horizon (CAH), $\R=\rA=a\,\mathcal{r}_{\text{A}}$. When the surface is required to define a global causal boundary, the radial function is fixed as $\mathcal{r}(t)=\mathcal{r}_{\text{E}}(t)$, and then Eq.~\eqref{ASB} reduces to the cosmological event horizon (CEH). The only difference between them is the constraint imposed on the radial function $\mathcal{r}(t)$. The explicit expressions for $\rA$, $\rE$, and their time evolution are presented in the following section. This demonstrates that both commonly studied cosmological horizons arise naturally as special cases within the
framework of arbitrary spherically symmetric surfaces.

 Substituting Eq.~\eqref{fluidT} into the conservation law~\eqref{CDTM} yields the continuity equation,
\begin{align}
\label{continuityequationformatter}
\dot\rho+3H(\rho+P)=0.
\end{align}
Multiplying both sides by $\tilde V dt$, where $\tilde V=4\pi \R^3/3$ is the volume enclosed by the spherical surface, gives
\begin{align}
\tilde{V} d\rho
+3H\tilde V(\rho+P)dt=0.
\end{align}
Adding $(\rho+P)d\tilde V$ to both sides and using
$
d\tilde V
=3\tilde V\frac{\dot{\tilde r}}{\tilde r}dt,
$
one obtains
\begin{equation}
\begin{aligned}
\label{heatingST}
dE+Pd\tilde V
&=
(\rho+P)
3\left(
\frac{\dot{\tilde r}}{\tilde r}-H
\right)\tilde Vdt
\\
&=
(\rho+P)\Delta,
\end{aligned}
\end{equation}
where $E=\rho\tilde V$ denotes the total matter energy enclosed by the spherical surface, and $\Delta \equiv 3\left(\frac{\dot{\tilde r}}{\tilde r}-H
\right)\tilde Vdt$. When evaluated on the CAH, $E$ reduces to the corresponding Misner-Sharp energy~\cite{PhysRev.136.B571}.

Eq.~\eqref{heatingST} describes the change of matter energy under mechanical work together with an additional contribution proportional to the term denoted by $\Delta$. Since the left-hand side possesses the familiar structure of first law of ordinary thermodynamics, the extra term on the right-hand side naturally calls for a thermodynamic interpretation. This provides the first indication that the spherical surface itself may carry thermodynamic significance.

To clarify the physical meaning of $\Delta$, consider first a strictly comoving surface, $\tilde r_{\text{c}}(t)=a(t)\mathcal{r}_{\text{c}}$, with constant $\mathcal{r}_{\text{c}}$. In this case,
$
\dot{\tilde r}_{\text{c}}/\tilde r_{\text{c}}=H,
$
and the right-hand side of Eq.~\eqref{heatingST} vanishes identically. One then recovers the standard first law for a closed adiabatic system,
$
dE+Pd\tilde V=0.
$
The situation changes when $\Delta\neq0$, which indicates  the presence of an additional contribution beyond the usual adiabatic evolution. In principle, this contribution may originate from heat conduction, gravitational work, or a genuine matter flux across the spherical surface. The first possibility can be ruled out immediately, since a perfect fluid does not support heat conduction. The second possibility is more subtle within the framework of general relativity. Although gravity is geometrized and exerts no local four-force on freely falling matter, notions of gravitational work can nevertheless arise in non-geodesic or radiative settings through quasi-local energy constructions~\cite{brown1993quasilocal,misner1973gravitation}. Since the cosmic fluid considered here follows geodesic motion, we shall not pursue this interpretation further. In these situations, it is natural to interpret $\Delta$ as characterizing the net flux of matter crossing the spherical surface. At any instant, the spherical surface may be compared with a comoving surface enclosing the same volume. If $\Delta<0$, the spherical surface expands more slowly than the corresponding comoving one, leading to a net outward flux of matter. Conversely, $\Delta>0$ implies that the surface expands more rapidly and hence acquires matter from its surroundings. Thus, the sign of $\Delta$ directly determines the direction of matter transport across the surface.

Let $n$ denote the particle number density. The change in particle number enclosed by the spherical surface during the interval $dt$ is
$dN=n\Delta\equiv
3n\left(
\frac{\dot{\tilde r}}{\tilde r}-H
\right)\tilde Vdt.$
Since $n>0$, the quantity denoted by $\Delta$ provides a direct measure of matter flux through the surface. We therefore define
\begin{equation}
\label{matterfluxmeasure}
\Delta \equiv 3\left(\frac{\dot{\tilde r}}{\tilde r}-H
\right)\tilde Vdt
\end{equation}
and refer to it as the matter-flux measure. A related quantity was obtained kinematically for the CAH in Ref.~\cite{PhysRevD.109.103515}. However, in the present framework it arises dynamically from the conservation law governing matter evolution.

The significance of $\Delta$ stems from the fact that it originates directly from matter dynamics while appearing in a relation with an explicit thermodynamic structure. Consequently, $\Delta$ admits a natural thermodynamic interpretation. Since it directly measures the matter flux through an arbitrary spherically symmetric surface, such surfaces naturally possess thermodynamic significance and can be used to characterize the cosmic thermodynamic properties. Furthermore, owing to the Friedmann equation, $\Delta$ can be expressed entirely in geometric terms. This establishes a link between thermodynamics and cosmic dynamics on arbitrary spherically symmetric surfaces. In this sense, it provides a manifestation of the deep connection between geometry and thermodynamics in gravitational systems beyond the conventional horizon framework.

\subsection{Clausius relation}
\label{sec:ClausiusASB}
In the previous subsection, we argued that arbitrary spherically symmetric surfaces can be assigned thermodynamic properties, much like cosmological horizons. We now proceed to derive the thermodynamic law associated with such surfaces, namely the Clausius relation.

Conventionally, the Clausius relation is regarded as the starting point for deriving gravitational field equations~\cite{Jacobson1995,Cai2005}. The analysis presented here suggests a different perspective. Specifically, we shall demonstrate that the Clausius relation arises naturally from gravitational dynamics. The derivation relies crucially on Eq.~\eqref{heatingST} and on the thermodynamic interpretation of the matter-flux measure $\Delta$ developed in the preceding subsection.

Since the area law of horizon entropy is firmly established within general relativity, it is natural, in the spirit of the holographic principle, to associate an entropy-like quantity with the spherical surface~\eqref{ASB}. We therefore postulate
\begin{align}
\mathcal{S}=\mathcal{C} A,
\label{entropylike}
\end{align}
where $A=4\pi\R^2$ denotes the area of the surface and $\mathcal{C}$ is an undetermined coefficient. At this stage, $\mathcal C$ is allowed to depend on both the cosmological evolution and the particular spherical surface under consideration, i.e. $\mathcal{C}=\mathcal{C}(t,\R,\dR)$. As we shall demonstrate below, the requirement of equilibrium-thermodynamics consistency uniquely fixes this undetermined coefficient, thereby recovering the standard Bekenstein area law.

Taking the differential of Eq.~\eqref{entropylike} and multiplying the resulting expression by $\R$, one obtains
 \begin{align}
    \tilde{r} d\mathcal{S} = 2\mathcal{C}d\tilde{V}+3\tilde{V}d\mathcal{C}.\label{rdS}
 \end{align}
 The appearance of the combination $(\rho+P)$ in Eq.~\eqref{heatingST} suggests making use of the second Friedmann equation~\eqref{F2}. Multiplying Eq.~\eqref{F2} by $\tilde{r} d\mathcal{S}/(2\pi)$  and subsequently employing both the continuity equation~\eqref{continuityequationformatter} and Eq.~\eqref{rdS}, one finds
 \begin{equation}
 \begin{aligned}
   \frac{\tilde{r}}{2\pi}\left(\dot{H}-\frac{k}{a^2} \right)d\mathcal{S} &=-4 G \mathcal{C}\left(dE+Pd\tilde{V} \right) -
     \frac{3}{2}\left(\rho +P \right)\tilde{V}4 Gd\mathcal{C}
     \\
     &-4G\mathcal{C}\tilde{V}\left[3H\left( \rho+P\right) \right]dt.
 \end{aligned}
 \end{equation}
 Employing Eq.~\eqref{F2} and the relation~\eqref{entropylike}, one finds that the last term in the above equation can be expressed as
 \begin{align}
 -4G\mathcal{C}\tilde{V}\left[3H\left( \rho+P\right) \right]dt=\frac{1}{2\pi}H\left(\dot{H}- \frac{k}{a^2}\right)\frac{\tilde{r}^2}{\dot{\tilde{r}}} d\mathcal{S}+ \frac{3}{2}\left(\rho +P \right)\tilde{V}H\frac{\tilde{r}}{\dot{\tilde{r}}}4 Gd\mathcal{C}.
 \end{align}
 Substituting this result back into the preceding equation and rearranging terms, one obtains the first thermodynamic-like relation
 \begin{align}
    \mathcal{T}_1d\mathcal{S} =-4 G \mathcal{C}\left(dE+Pd\tilde{V} \right) +
     \frac{3}{2}\left(\rho +P \right)\tilde{V}\left(\frac{H}{\dot{\tilde{r}}/\tilde{r}} -1\right)4 Gd\mathcal{C},\label{T1GR}
 \end{align}
 where
 \begin{align}
 \label{Temperature1}
 \mathcal{T}_1 =\frac{1}{2\pi}\left[\left(\dot H - \frac{k}{a^2} \right)\tilde{r}-H\left(\dot H- \frac{k}{a^2}\right)\frac{\tilde{r}^2}{\dot{\tilde r}} \right].
 \end{align}

By employing the explicit form of the temperature $\mathcal{T}_1$ and the Friedmann equation~\eqref{F2}, the term proportional to $d\mathcal{C}$ in Eq.~\eqref{T1GR} can be expressed entirely in terms of geometric quantities. After some straightforward algebra, one obtains the exact identity as
\begin{equation}
\frac{3}{2}\left(\rho +P \right)\tilde{V}\left(\frac{H}{\dot{\tilde{r}}/\tilde{r}} -1\right)4 Gd\mathcal{C} = \mathcal{T}_1 A\,d\mathcal{C}.
\label{dCidentity}
\end{equation}
Substituting Eq.~\eqref{dCidentity} into~\eqref{T1GR} yields
\begin{align}
\mathcal{T}_1 d\mathcal{S} = -4 G \mathcal{C}\left(dE+Pd\tilde{V} \right) + \mathcal{T}_1 A\,d\mathcal{C}.
\label{T1GRTAdC}
\end{align}
This reformulation is physically illuminating. Within the framework of non-equilibrium thermodynamics~\cite{deGrootMazur1984,Prigogine1967}, the total entropy variation can be decomposed as $d\mathcal{S} = d_e\mathcal{S} + d_i\mathcal{S}$, where $d_e\mathcal{S}$  denotes the entropy flow associated with energy exchange with the surroundings, while $d_i\mathcal{S} \ge 0$ represents the entropy generated by irreversible processes within the system. Multiplying this decomposition by $\mathcal{T}_1$ and comparing it with Eq.~\eqref{T1GRTAdC}, one is naturally led to identify the term $\mathcal{T}_1 A\,d\mathcal{C}$ as the contribution arising from internal entropy production. If the coefficient $\mathcal{C}$ varies during cosmic evolution, the spherical surface behaves as a non-equilibrium thermodynamic system. Conversely, requiring that the system admit a consistent equilibrium thermodynamic description imposes  $d\mathcal{C}=0$, implying that $\mathcal{C}$ must be constant. Eq.~\eqref{T1GRTAdC} then reduces to
\begin{equation}
\mathcal{T}_1\, d\mathcal{S} = -4 G \mathcal{C}(dE+Pd\tilde{V}).
\end{equation}

To render the physical content of this relation manifest, it is convenient to normalize the overall constant factor through a rescaling of units. Without loss of generality, one may choose
\begin{equation}
4G\mathcal{C}=1
\end{equation}
which immediately yields
\begin{equation}
\label{S=A/4G}
\mathcal{S} = \mathcal{C}A = \frac{A}{4G}.
\end{equation}
It is noteworthy that the Bekenstein entropy is not assumed a priori in the present framework. Instead, it emerges as a consequence of thermodynamic consistency, providing further evidence for the deep connection between gravity and thermodynamics.

 With $\mathcal{C}=1/4G$, Eq.~\eqref{T1GR} reduces to
 \begin{align}
 \mathcal{T}_1d\mathcal{S} = - \left(dE+Pd\tilde{V} \right)\label{T1}.
 \end{align}
 The above relation bears a strong resemblance to the conventional first law of thermodynamics, differing only by an overall minus sign. One might attempt to absorb this minus sign into the definition of the temperature-like quantity  $\mathcal{T}_1$  so as to recover the standard first-law form. However, such a redefinition fails to reproduce the correct work density when the present formalism is specialized to the cosmological apparent horizon. Consequently, Eq.~\eqref{T1} cannot be identified with the first law of thermodynamics.

 A more revealing result emerges from combining Eq.~\eqref{T1} with Eq.~\eqref{heatingST}, which yields
 \begin{equation}
\mathcal{T}_1\dot{\mathcal{S}}dt =-\left(\rho + P\right)\Delta=-3\left(\rho + P\right)\left(\frac{\dot{\tilde r}}{\tilde r}-H \right)\tilde V dt
\end{equation}
 Using Eqs.~\eqref{Temperature1} and~\eqref{S=A/4G}, the expansion of the above equation is given by
 \begin{equation}
\frac{1}{2\pi}\left[\left(\dot H - \frac{k}{a^2} \right)\tilde{r}\frac{d\mathcal{S}}{d\R}\dR+H\left(\frac{k}{a^2}-\dot H\right)\frac{\tilde{r}^2}{\dot{\tilde r}} \frac{d\mathcal{S}}{d\R}\dR\right]dt =-3\left(\rho + P\right)\left(\frac{\dot{\tilde r}}{\tilde r}-H \right)\tilde V dt
\end{equation}
 Taking the limit $\dR \to 0$, the above equation reduces to
\begin{equation}
\label{PreClausius}
\frac{1}{2\pi}H\left(\frac{k}{a^2}-\dot{H}\right)\R^2\frac{d\mathcal{S}}{d\R}dt=3H\left(\rho +P \right)\tilde{V}dt
\end{equation}
 which takes a form similar to the Clausius relation from which Jacobson derived the Einstein field equations~\cite{Jacobson1995}. To make this thermodynamic structure explicit, we introduce the so-called Jacobson temperature:
 \begin{align}
 \label{TJ}
 T_{\text{J}} = \frac{1}{2\pi}H\left(\frac{k}{a^2}-\dot{H}\right)\frac{\tilde{r}^2}{\dot{\tilde r}}
 \end{align}
 Then Eq.~\eqref{PreClausius} can be written as
 \begin{align}
 T_{\text{J}}\delta \mathcal{S}=\delta Q\label{ClausiusASB},
 \end{align}
 where the heat flux across the spherical surface with radius $\R$ is given by
 \begin{equation}
\label{heatfluxASB}
\delta Q = 3H\left(\rho +P \right)\tilde{V}dt.
\end{equation}
 The symbol ``$\delta$'' in the above expressions indicates that the limit $\dR \to 0$ is taken, corresponding to a quasi-static process or virtual displacement consistent with the instantaneously stationary assumption adopted in the original work of Jacobson~\cite{Jacobson1995}. This is why no work term appears in Eq.~\eqref{ClausiusASB}.

 Since the Jacobson temperature $T_{\text{J}}$ reduces to the Cai-Kim temperature when evaluated on the CAH (see Sec.~\ref{sec:TAH} for details), and the heat flux~\eqref{heatfluxASB} generalizes the one across the CAH, which can be derived from Jacobson's original formalism using the Kodama vector~\cite{Cao2010}, we arrive at three main conclusions. First, arbitrary spherically symmetric surfaces are naturally endowed with the standard Bekenstein entropy. Second, they possess a well-defined temperature determined entirely by geometric quantities. Third, the Clausius relation on such surfaces emerges directly from gravitational dynamics. In this sense, the Clausius relation should be regarded not as a fundamental postulate, but as a consequence of gravity itself.

\subsection{Unified first law of thermodynamics}
Once the actual evolution of a dynamical surface and the work performed by the cosmic medium are taken into account, a more complete thermodynamic relation is required.

To recover the correct work density, $W=\frac{1}{2}(\rho-P)$, and thereby obtain the unified first law, we add the term $-\frac{1}{4\pi}\left(\dot H -\frac{k}{a^2} \right)\tilde{r}d\mathcal{S}$ to both sides of Eq.~\eqref{T1GR}. Note that this term can be rewritten as
\begin{align}
-\frac{1}{4\pi}\left(\dot H -\frac{k}{a^2} \right)\tilde{r}d\mathcal{S}=4G\mathcal{C}\frac{1}{2}\left(\rho+P\right)d\tilde{V} + \frac{3}{4}\left(\rho+P \right)\tilde{V}4G\mathcal{C}d\mathcal{C}
\end{align}
where Eqs.~\eqref{F2} and~\eqref{rdS} have been employed. After straightforward algebra, one thus arrives at the second thermodynamic-like relation:
 \begin{equation}
 \begin{aligned}
 \label{T2GR}
 \frac{1}{2\pi}\left[\frac{1}{2}\left(\dot H - \frac{k}{a^2} \right)\tilde{r}-H\left(\dot H- \frac{k}{a^2}\right)\frac{\tilde{r}^2}{\dot{\tilde r}} \right]
d\mathcal{S} &=4G\mathcal{C}\left(-dE+Wd\tilde V\right)
\\
&+ \frac{3}{4}\left(\rho +P \right)\tilde{V}\left(\frac{2H}{\dot{\tilde{r}}/\tilde{r}} -1\right)4 Gd\mathcal{C},
 \end{aligned}
 \end{equation}
where $W= \frac{1}{2}\left(\rho-P\right)$ is the work density. As in the derivation of the Clausius relation, thermodynamic consistency uniquely fixes the coefficient to $\mathcal{C}=1/4G$. Consequently, Eq.~\eqref{T2GR} reduces to
\begin{equation}
 \label{UFLASB}
\mathcal{T}_2 d\mathcal{S} =-dE+Wd\tilde V,
\end{equation}
where
\begin{equation}
\label{T2}
\mathcal{T}_2 \equiv \frac{1}{2\pi}\left[\frac{1}{2}\left(\dot H - \frac{k}{a^2} \right)\tilde{r}-H\left(\dot H- \frac{k}{a^2}\right)\frac{\tilde{r}^2}{\dot{\tilde r}} \right].
\end{equation}

Equation~\eqref{UFLASB} can be identified as the unified first law for arbitrary spherically symmetric surfaces. Unlike its apparent-horizon counterpart, it does not emerge as a purely geometrical identity. Instead, it encapsulates the dynamical interplay between matter and spacetime and provides a thermodynamic description of the evolution of the surface.

\section{Applications}
\label{sec:APP}
As shown in the previous section, arbitrary spherically symmetric surfaces in non-flat FLRW spacetimes naturally exhibit well-defined thermodynamic properties, including the Clausius relation and the unified first law of thermodynamics. In this section, we show that these general results consistently recover the established thermodynamic description of the CAH. We further demonstrate that applying the general framework to the CEH yields a self-consistent thermodynamic formulation, thereby resolving the long-standing ambiguities in the literature concerning both its temperature and the corresponding thermodynamic laws.

\subsection{Cosmological apparent horizon}
\label{sec:TAH}

 The CAH is a locally defined trapping horizon, characterized by the vanishing expansion of the ingoing null geodesic congruence, $\theta_{n}=0$, while the expansion of the outgoing null geodesic congruence remains positive, $\theta_{l}>0$~\cite{faraoni_cosmological_2015}.  This condition fixes the radial coordinate and the proper radius of the CAH to be
\begin{align}
\label{loctionAH}
\mathcal{r}_{\text{A}} = \frac{1}{\sqrt{\dot a^2+k}},
\end{align}
and
\begin{align}
\label{RadiusAH}
\rA \equiv a\mathcal{r}_{\text{A}} = \frac{1}{\sqrt{H^2 +k/a^2}},
\end{align}
respectively. The time evolution of the proper radius is obtained directly as
\begin{equation}
\label{drA}
\drA = -H\rA^3\left(\dot{H}-\frac{k}{a^2}\right).
\end{equation}
Although the CAH is, in general, a non-null surface, its local geometric definition together with its spherical symmetry makes it a natural thermodynamic boundary.

To verify that our general formalism consistently reproduces the standard thermodynamics of the CAH, we first evaluate the matter-flux measure defined in Eq.~\eqref{matterfluxmeasure}. Using Eq.~\eqref{drA}, the relative expansion rate is
\begin{align}
\frac{\dot{\tilde r}_{\text{A}}}{\tilde r_{\text{A}}} - H  =H\left( \frac{-(\dot H - k/a^2)}{H^2 + k/a^2}-1\right).
\end{align}
Substituting this expression into Eq.~\eqref{matterfluxmeasure} gives
\begin{align}
\label{OPENSYSAH}
\Delta\vert_{\tilde{r}=\rA}=3V_{\text{A}}H\left[\frac{-(\dot{H}-k/a^2)}{H^2+k/a^2}-1 \right]dt,
\end{align}
where $V_{\text{A}} = 4\pi \rA^3/3$ denotes the proper volume enclosed by the apparent horizon. In the spatially flat case ($k=0$), this expression reduces to $\Delta\vert_{\tilde{r}=\rA}= \frac{4\pi}{H^2}\left(-\dot H/H^2 -1 \right)dt$, which agrees exactly with the result reported in Ref.~\cite{PhysRevD.109.103515}. Furthermore, adopting the equation of state $P=w\rho$, Eq.~\eqref{OPENSYSAH} can be rewritten as
\begin{align}
\Delta\vert_{\rA}=\frac{3}{2}V_{\text{A}}H\left(1+3w \right)dt.
\end{align}
This expression establishes a direct connection between the matter-flux across the CAH and the cosmic equation of state. When the strong energy condition is satisfied ($w>-1/3$ with $\rho>0$), one has $\Delta>0$. It implies a net influx of matter into the CAH during a decelerating phase of cosmic expansion. By contrast, for an accelerating universe with $w<-1/3$, the strong energy condition is violated and $\Delta<0$, which corresponds to a net outward matter flux. The persistence of a non-vanishing matter flux therefore demonstrates that the CAH constitutes an open thermodynamic system, whose entropy evolution must incorporate the exchange of matter across the horizon~\cite{PhysRevD.109.103515,PhysRevD.111.043544}.

We next evaluate the general thermodynamic quantities on the CAH. The entropy defined by Eq.~\eqref{S=A/4G} becomes
\begin{equation}
S^{\text{A}} = \mathcal{S}\vert_{\R=\rA} = \frac{\pi \rA^2}{G}.
\end{equation}
This entropy-area relation, which follows directly from the general framework, serves as the starting point for the thermodynamic study of the cosmological apparent horizon.
The Jacobson temperature~\eqref{TJ} and the heat flux~\eqref{heatfluxASB} specialize to
\begin{equation}
\begin{aligned}
T_{\text{J}}^{\text{A}} &= T_{\text{J}}\vert_{\R=\rA} = \frac{1}{2\pi}H\left(\frac{k}{a^2}-\dot{H}\right)\frac{\rA^2}{\drA}, \\
\delta Q^{\text{A}} &= \delta Q\vert_{\R=\rA} = 3H(\rho +P)V_{\text{A}}dt,
\end{aligned}
\end{equation}
where $V_{\mathrm{A}}=4\pi\rA^{3}/3$ is the volume enclosed by the apparent horizon. Substituting Eq.~\eqref{drA} into the expression for $T_{\mathrm{J}}^{\mathrm{A}}$ gives
\begin{align}
T_{\text{J}}^{\text{A}} = \frac{1}{2\pi}H\left(\frac{k}{a^2}-\dot{H}\right)\frac{\rA^2}{-H\rA^3(\dot{H}-k/a^2)} = \frac{1}{2\pi \rA},
\end{align}
which is independent of $\drA$. Consequently, these quantities satisfy the Clausius relation exactly,
\begin{equation}
T_{\text{J}}^{\text{A}}\delta S^{\text{A}}=\delta Q^{\text{A}}.
\end{equation}
Although the derivation of the Clausius relation in Sec.~\ref{sec:ClausiusASB} relied on the particular limit $\dR \to 0$, the above result remains exact because the special form of Eq.~\eqref{drA} eliminates the dependence of $T_{\text{J}}^{\text{A}}$ on $\drA$. Consequently, the Clausius relation holds on the physical CAH without invoking any approximation.

 Likewise, evaluating the general temperature $\mathcal{T}_2$ defined in Eq.~\eqref{T2} on the CAH gives the Kodama--Hayward temperature:
\begin{equation}
\begin{aligned}
\label{TAH}
T^{\text{A}}_{\text{KH}} = \mathcal{T}_{2}\vert_{\R=\rA}
&= \frac{1}{2\pi}\left[\frac{1}{2}\left(\dot H - \frac{k}{a^2}\right)\rA - H\left(\dot H- \frac{k}{a^2}\right)\frac{\rA^2}{\drA} \right]
\\
&= \frac{1}{2\pi\rA}\left(1-\frac{\drA}{2H\rA} \right)
\\
&= -\frac{\kappa}{2\pi},
\end{aligned}
\end{equation}
where $\kappa$ denotes the surface gravity of the CAH~\cite{Cao2010}. Substituting this temperature together with the work density $W=(\rho-P)/2$ into Eq.~\eqref{UFLASB} immediately yields
\begin{equation}
\label{UFLAH}
T^{\text{A}}_{\text{KH}}dS^{\text{A}} = -dE_{\text{A}} + WdV_{\text{A}},
\end{equation}
where $E_{\text{A}} = \rho V_{\text{A}}$ is the Misner-Sharp energy enclosed by the apparent horizon. Equation~\eqref{UFLAH} is precisely the standard unified first law associated with the CAH~\cite{Cai2005,AkbarCai2007,CaiCao2007}. Hence, the general thermodynamic framework developed here consistently reproduces the established thermodynamic description of the apparent horizon.

 When the continuity equation~\eqref{continuityequationformatter} and the Friedmann equation~\eqref{F2} are taken, the thermodynamic relations~\eqref{T1} and~\eqref{UFLASB} become identical upon adopting the Jacobson temperature $T_{\text{J}}$,
\begin{equation}
 T_{\text{J}}d\mathcal{S} = 3H\tilde{V}(\rho +P)dt.
\end{equation}
Specializing to the CAH, where $T_{\text{J}}\vert_{\R=\rA} = T_{\text{CK}}^{\text{A}} = 1/(2\pi \rA) > 0$, one obtains $T_{\text{CK}}^{\text{A}} dS^{\text{A}} = 3H V_{\text{A}} (\rho+P)dt$. Since the Cai--Kim temperature is strictly positive, the sign of $dS^{\text{A}}$ is determined solely by $(\rho+P)$. Therefore, for any cosmic fluid satisfying the null energy condition ($\rho+P>0$, or equivalently $w>-1$), one has $dS^{\text{A}}>0$. This implies that the entropy of the apparent horizon is non-decreasing.

\subsection{Cosmological event horizon}
\label{sec:TEH}

 Besides the CAH, another cosmological horizon of fundamental interest is the CEH. Its existence requires an accelerating universe, a scenario that is now firmly established by cosmological observations~\cite{Riess_1998,Perlmutter_1999}. In the following, we specialize our general framework to the CEH without introducing any assumptions beyond spherical symmetry.

 The CEH is a global null surface defined as the causal boundary separating events that can ever be observed by a comoving observer located at the origin, $\mathcal{r}=0$. Its coordinate position, $\mathcal{r}_{\text{E}}(t)$, is determined by integrating the ingoing radial null geodesic,
 \begin{align}
 \label{loctionEH}
 \int_t^{\infty}\frac{dt'}{a(t')}=-\int_{\mathcal{r}_{\text{E}}(t)}^0\frac{dr}{\sqrt{1-kr^2}}.
 \end{align}
The corresponding proper radius is $\rE=a\mathcal{r}_{\text{E}}$. Differentiating Eq.~\eqref{loctionEH} with respect to cosmic time and using the chain rule $d/dt = \dot{\mathcal{r}}_{\text{E}}d/d\mathcal{r}_{\text{E}}$ give the evolution equation:
 \begin{align}
 \label{dotEH}
 \dot{\tilde {r}}_{\text{E}}= H\rE - \sqrt{1-\frac{k\rE^2}{a^2}}.
 \end{align}
 For a spatially flat universe ($k=0$), this reduces to $\dot{\tilde {r}}_{\text{E}} = H\rE-1$, consistent with the one given in Ref.~\cite{PhysRevD.84.024003}.

 Since the CEH is a null causal boundary, it is natural to associate an entropy with it following Bekenstein's original argument. Like a black hole event horizon, the CEH acts as a one-way causal membrane: once matter crosses the horizon, it irreversibly leaves the causal past of the comoving observer, together with the information it carries. This irreversible loss of accessible information provides the physical basis for attributing an intrinsic entropy to the CEH, in close analogy with black-hole thermodynamics. Within our general framework, this interpretation arises naturally.

Substituting the evolution equation~\eqref{dotEH} into Eq.~\eqref{matterfluxmeasure}, we obtain the matter-flux measure on the CEH as
\begin{align}
\label{OPENSYSEH}
\Delta\vert_{\tilde{r}=\rE} = 3V_{\text{E}}\left(\frac{\dot{\tilde r}_{\text{E}}}{\tilde r_{\text{E}}}-H\right)dt = -3V_{\text{E}}\frac{\sqrt{1-\frac{k\rE^2}{a^2}}}{\rE}dt < 0,
\end{align}
where $V_{\text{E}} = 4\pi \rE^3/3$ is the proper volume enclosed by the CEH. In contrast to the CAH, the matter flux across the CEH is strictly negative. This signifies a perpetual outflow of matter, consistent with the global causal structure of the event horizon. The persistent outward matter flux reflects the one-way causal nature of the CEH and provides a thermodynamic basis for attributing entropy to the horizon. Indeed, to preserve the generalized second law, the entropy carried away by matter must be compensated by an increase in the horizon entropy.

Earlier attempts to construct consistent thermodynamics for the CEH using Jacobson's approach encountered difficulties, largely because they relied on ad hoc definitions of temperature~\cite{wang_thermodynamics_2006,chakraborty_is_2012,tu_general_2016}. In contrast, our formalism, which falls within the gravity-to-thermodynamics paradigm, derives both the entropy and the temperature from first principles without any additional assumptions.

Having established the thermodynamic role of the CEH through its matter flux, we next evaluate the corresponding thermodynamic quantities. According to Eqs.~\eqref{S=A/4G},~\eqref{TJ} and~\eqref{heatfluxASB}, the entropy, Jacobson temperature, and heat flux are given by
\begin{equation}
\begin{aligned}
S^{\text{E}} &= \mathcal{S}\vert_{\R=\rE} = \frac{\pi \rE^2}{G}, \\
T_{\text{J}}^{\text{E}} &= T_{\text{J}}\vert_{\R=\rE} = \frac{1}{2\pi}H\left(\frac{k}{a^2}-\dot{H}\right)\frac{\rE^2}{\drE}, \\
\delta Q^{\text{E}} &= \delta Q\vert_{\R=\rE} = 3H(\rho +P)V_{\text{E}}dt.
\end{aligned}
\end{equation}
These expressions coincide with those priori postulated in Ref.~\cite{tu_general_2016}, but are obtained here directly from the general framework developed in previous section. Employing the Friedmann equations together with Eq.~\eqref{dotEH}, one finds the exact identity,
\begin{equation}
\label{ClausisusEH}
T_{\text{J}}^{\text{E}}\,\delta S^{\text{E}} = \delta Q^{\text{E}},
\end{equation}
which is precisely the Clausius relation on the CEH.

The unified first law of thermodynamics follows by evaluating Eq.~\eqref{UFLASB} on the CEH, i.e.,
\begin{align}
\label{UFLEH}
T^{\text{E}}dS^{\text{E}} = -dE_{\text{E}} + WdV_{\text{E}},
\end{align}
where the temperature $T^{\text{E}}$ is
\begin{align}
\label{TEH}
T^{\text{E}}= \mathcal{T}_{2}\vert_{\R=\rE}
= \frac{1}{2\pi}\left[\frac{1}{2}\left(\dot H - \frac{k}{a^2} \right)\rE - H\left(\dot H- \frac{k}{a^2}\right)\frac{\rE^2}{\drE} \right].
\end{align}

As a non-trivial consistency check, we verify that this temperature reduces to the well-known de Sitter temperature in the appropriate limit. In a pure de Sitter universe ($k=0$, $\dot H = 0$), the event horizon coincides with the Hubble radius, $\rE = H^{-1}$, and its evolution equation~\eqref{dotEH} gives $\drE = H\rE - 1 = 0$.  Substituting these relations into Eq.~\eqref{TEH}, the first term vanishes identically, whereas the second term requires a careful limiting procedure. Using $\drA = -H\rA^3(\dot H - k/a^2)$ the second term becomes $(\rE^2\drA)/(2\pi\rA^3\drE)$. Since the CEH and CAH coincide in the de Sitter limit, $\drA/\drE \to 1$ and one thus has
\begin{equation}
\lim_{\substack{\dot{H} \to 0 \\ k \to 0}} T^{\text{E}} = \frac{H}{2\pi},
\end{equation}
which is precisely the Gibbons--Hawking temperature~\cite{gibbons_cosmological_1977}. Thus, the general temperature derived here correctly reproduces the Gibbons-Hawking temperature in the de Sitter limit.

Previous attempts to formulate the thermodynamics of the CEH within Jacobson's approach were hindered by the absence of a uniquely defined horizon temperature and therefore relied on phenomenological prescriptions~\cite{wang_thermodynamics_2006,chakraborty_is_2012,tu_general_2016}. By contrast, our gravity-to-thermodynamics framework determines both the entropy and the temperature directly from the underlying gravitational dynamics, without introducing any additional assumptions or ad hoc inputs.

As an illustration, we consider a flat universe ($k=0$) filled with a non-interacting mixture of holographic dark energy (HDE) and pressureless dust~\cite{chakraborty_is_2012}. The energy density of HDE is given by
\begin{align}
\rho_{\text{D}} = \frac{3c^2}{\rE^2}, \label{HDE}
\end{align}
where $c$ is a dimensionless constant. The relevant Friedmann equations are
\begin{align}
H^2 = \frac{1}{3}\left(\rho_{\text{m}}+\rho_{\text{D}} \right),\quad \dot{H} = -\frac{1}{2}\left(\rho_{\text{m}}+\rho_{\text{D}}+P_{\text{D}} \right), \label{F1F2HDE}
\end{align}
Here we work in units where $8\pi G =1$, $\rho_{\text{m}}$ is the energy density of pressureless dust and the pressure of HDE is $P_{\text{D}} = w_{\text{D}}\rho_{\text{D}}$, where the equation-of-state parameter $w_{\text{D}}$ is~\cite{WANG2005141}
\begin{align}
w_{\text{D}} = -\frac{1}{3} - \frac{2\sqrt{\Omega_{\text{D}}}}{3c},
\end{align}
where $\Omega_{\text{D}} = \rho_{\text{D}}/(3H^2)$ is the density parameter of the HDE. These relations imply that $\rE^{-1}=\sqrt{\rho_{\text{D}}/(3c^2)}=\frac{H}{c}\sqrt{\Omega_{\text{D}}}$. Substituting this relation into Eq.~\eqref{dotEH} gives
\begin{align}
\frac{\drE}{\rE} = H - \frac{1}{\rE} = H\left(1 - \frac{\sqrt{\Omega_{\text{D}}}}{c} \right),
\end{align}
which agrees with the one given in Ref.~\cite{wang_thermodynamics_2006}. The corresponding heat flux across the CEH is then
\begin{equation}
\begin{aligned}
\delta Q^{\text{E}} &= 3V_{\text{E}}H\left( \rho_{\text{m}}+\rho_{\text{D}}+P_{\text{D}}\right)dt \\
&= \frac{\rE^3}{2G}H^3\left(1 +  w_{\text{D}}\Omega_{\text{D}} \right)dt,
\end{aligned}
\end{equation}
where we have used Eq.~\eqref{F1F2HDE} together with the relation $\Omega_{\text{D}} = \rho_{\text{D}}/(3H^2)$ in calculations. For a spatially flat universe ($k=0$), the Jacobson temperature and the differential of the entropy become
\begin{equation}
\begin{aligned}
T_{\text{J}}^{\text{E}} &= \frac{1}{2\pi}H\left(-\dot{H}\right)\frac{\rE^2}{\drE}
= \frac{3H^2\rE}{4\pi \left(1-\sqrt{\Omega_{\text{D}}}/c \right)}\left(1 + w_{\text{D}}\Omega_{\text{D}} \right), \\
dS^{\text{E}} &= d\left(\frac{\pi \rE^2}{G} \right) = \frac{2\pi \rE^2}{G}H\left(1-\frac{\sqrt{\Omega_{\text{D}}}}{c} \right)dt.
\end{aligned}
\end{equation}
One readily verifies that the Clausius relation, $T_{\text{J}}^{\text{E}} d S^{\text{E}} = \delta Q^{\text{E}}$, is satisfied exactly, thus confirming Eq.~\eqref{ClausisusEH} for this non-trivial cosmological model. Consequently, the unified first law~\eqref{UFLEH} is likewise fulfilled. This example indicates that the inconsistencies reported in previous studies originate not from thermodynamics of the CEH itself, but from ad hoc and mutually incompatible prescriptions for the horizon temperature and entropy. By deriving these quantities directly from the underlying gravitational dynamics, the present framework naturally yields a self-consistent thermodynamic description of the CEH.

\section{Conclusion}
\label{sec:Conclusion}

In this work, we have systematically investigated the thermodynamics of arbitrary spherically symmetric surfaces in non-flat FLRW spacetimes within the framework of general relativity. Instead of postulating thermodynamic relations on horizons, we adopted the gravity-to-thermodynamics perspective, in which the thermodynamic laws follow directly from the gravitational field equations and the dynamics of matter. This framework eliminates the need for ad hoc assumptions about the temperature and entropy of spherical surfaces. It also extends the gravity-thermodynamics correspondence from causal horizons to arbitrary spherically symmetric surfaces. Furthermore, it naturally distinguishes equilibrium from non-equilibrium contributions and identifies the Bekenstein entropy as the unique entropy compatible with thermal equilibrium.

A key quantity introduced in this work is the matter-flux measure $\Delta$, which follows directly from the covariant conservation law of the matter energy-momentum tensor. This quantity characterizes the net matter flux across an arbitrary spherical surface and establishes a direct connection between matter dynamics and thermodynamics. Its appearance in an equation with an explicit first-law structure provides a fundamental justification for assigning thermodynamic properties to general spherical surfaces. As a result, the thermodynamic interpretation of gravity is no longer restricted to causal horizons but applies to arbitrary spherically symmetric surfaces. Within this framework, we derived two thermodynamic laws that hold on an arbitrary spherically symmetric surface: the Clausius relation associated with the temperature $T_{\mathrm{J}}$ and the unified first law associated with the temperature $\mathcal{T}_2$. Although both laws arise directly from the gravitational dynamics, they describe different physical processes. The Clausius relation is formulated for a virtual displacement of the surface, or equivalently under a quasi-static approximation, and therefore contains no work term. By contrast, the unified first law describes the actual evolution of the surface and naturally incorporates the work term.

Applying the formalism to the cosmological apparent horizon, we recovered the standard Bekenstein entropy together with the Cai-Kim and Kodama-Hayward temperatures, as well as the corresponding Clausius relation and unified first law, without introducing any additional assumptions. This agreement with the established results provides a non-trivial validation of the consistency and general applicability of the present framework.

We then applied the formalism to the cosmological event horizon, whose thermodynamic description has long remained elusive owing to the absence of a universally accepted temperature definition. Previous studies typically relied on modified temperature prescriptions whose physical origin was unclear and whose applicability was often limited. In contrast, the present framework determines the event-horizon temperatures directly from the Friedmann equations and the horizon evolution. Specifically, evaluating the general expressions~\eqref{TJ} and~\eqref{T2} on the event horizon yields the self-consistent temperatures $T_{\mathrm{J}}\vert_{\R=\rE}$ and $\mathcal{T}_2\vert_{\R=\rE}$, without introducing any additional parameters or phenomenological rescalings. These temperatures satisfy both the Clausius relation and the unified first law exactly. In the de Sitter limit, they reduce to the Gibbons-Hawking temperature, $T_{\mathrm{GH}}=H/(2\pi)$, which provides an important consistency check on the formalism. We further validated this thermodynamic description by considering a model with holographic dark energy and pressureless dust. The analysis shows that the inconsistencies reported in previous studies arise from incompatible phenomenological temperature prescriptions rather than from any intrinsic deficiency in the thermodynamics of the cosmological event horizon.

The framework developed in this work provides a general thermodynamic description of arbitrary spherically symmetric surfaces in dynamical spacetimes. The thermodynamic descriptions of the cosmological apparent horizon and the cosmological event horizon are both recovered as special cases of the general formalism. This establishes a unified framework for the thermodynamics of both horizons and general spherical surfaces. These results support the view that the gravity-thermodynamics correspondence is a generic consequence of gravitational dynamics rather than a property unique to causal horizons.

\section*{Acknowledgements}

This work was supported by the National Natural Science Foundation of China under Grant No.12165021, the China Scholarship Council (CSC) and the Young Talent Programme under the Xingdian Talent Support Plan.

\bibliographystyle{apsrev}
\bibliography{TASSref}

\end{document}